\documentclass[twocolumn,aps,pra,groupedaddress,amsmath,amssymb,showpacs]{revtex4-1}
\usepackage{graphicx}
\usepackage{graphicx,epsfig,float,epstopdf}
\usepackage{bm}
\usepackage[colorlinks,	linkcolor=blue,	anchorcolor=blue,citecolor=blue,bookmarksnumbered]{hyperref}

\begin{document}
\preprint{}

\title{Quantum Reflections of Nonlocal Optical Solitons
in a Cold Rydberg Atomic Gas}
\author{Zhengyang Bai$^{1, 2}$, Qi Zhang$^{1}$, and Guoxiang Huang$^{1,3,4,*}$ }
\affiliation{$^1$State Key Laboratory of Precision Spectroscopy,
             East China Normal University, Shanghai 200062, China\\
             $^2$School of Physics and Astronomy, University of Nottingham, Nottingham, NG7 2RD, UK\\
             $^3$NYU-ECNU Joint Institute of Physics, New York University at Shanghai, Shanghai 200062, China\\
             $^4$Collaborative Innovation Center of Extreme Optics, Shanxi University, Taiyuan 030006, China
\\
             $^*$gxhuang@phy.ecnu.edu.cn}
\date{\today}

\begin{abstract}
Quantum reflection refers to a non-vanishing reflection probability in the absence of a classically turning point. Much attention has been paid to such reflections due to their fundamental, intriguing physics and potential practical applications. Here we propose a scheme to realize a quantum reflection of nonlocal nonlinear optical beams in a cold Rydberg atomic gas via electromagnetically induced transparency working in a dispersion regime. Based on the long-range interaction between Rydberg atoms, we found that the system supports low-power nonlocal optical solitons. Such nonlocal solitons can display a sharp transition between reflection, trapping, and transmission when scattered by a linear attractive potential, created by gate photons stored in another Rydberg state. Different from conventional physical systems explored up to now, the quantum reflection of the nonlocal optical solitons in the Rydberg atomic gas exhibits interesting anomalous behaviors, which can be actively manipulated by tuning the incident velocity and intensity of the probe field, as well as the nonlocality of the Kerr nonlinearity inherent in the Rydberg atomic gas.
The results reported here are not only useful for developing Rydberg nonlinear optics but also helpful for characterizing the physical property of the Rydberg gas and for designing novel nonlinear optical devices.


\end{abstract}

\maketitle

\section{Introduction}\label{sec1}

Quantum reflection (QR) is a classically forbidden reflection in which a microscopic particle reflects from a potential without reaching a classically turning point, a typical and direct consequence of the wave nature of the microscopic particle~\cite{Friedrich,Goodman}. In particular, QR can occur for microscopic particles when they are reflected by attractive potential wells. In the past decades, tremendous efforts were focused on the study on the QR of atoms and molecules hitting solid surfaces~\cite{Berkhout,Shimizu,Pasquini,Scott,Pasquini1,ZhaoBS,Stickler,Barnea}. Comparing with conventional particles, the QR of matter-wave solitons have low scattering loss and large reflection probability~\cite{Lee,LiuM,cornish,Ernst,Marchant,Marchant2}, which are helpful not only for deepening the fundamental understanding of quantum theory and but also for realizing many practical applications.

In recent years, considerable efforts have been devoted to the investigation on interfacing light with strongly interacting Rydberg atomic gases under conditions of the electromagnetically induced transparency (EIT) (see the reviews given by Refs.~\cite{Saffman,Pri,Fir,Mur} for details).  One of main motivations for such investigation is due to the fact that Rydberg states have long coherent lifetime and extremely strong interaction (i.e. Rydberg-Rydberg interaction) between remote atoms, which can be effectively mapped to strong photon-photon interaction~\cite{Gor0,Busche}. As a result, giant Kerr nonlinearity at very low and even single photon level can be realized, which can be many orders of magnitude larger than that obtained via conventional optical media~\cite{pritchard2,Sev,Gor0,Stan,First,LLi,bing,Busche,Gullans,Liang,Bai,
ZhangQi,Hang,Bai2019,Hang2} and may be actively controlled by tuning system parameters. Recently, many experiments have demonstrated such strong, controllable photon-photon interaction~\cite{Busche,First,Liang} in Rydberg atomic gases, which can be utilized to design a broad range of novel optical devices, such as single-photon switches
and transistors~\cite{Baur,Tiarks,Gorn1}, quantum phase gates~\cite{Tiarks2,Tiarks3}, and deterministic single-photon sources~\cite{Ripka}, and so on.

Owing to the fact that the Maxwell's wave equation for the electric field in electrodynamics under a paraxial approximation is mathematically equivalent to the Schr\"{o}dinger equation in quantum mechanics, it is natural and will be interesting to extend the QR study beyond matter waves. In this work, we propose and analyze a scheme to realize an optical analogue of the QR of nonlocal nonlinear optical beams via a Rydberg-EIT working in a dispersion regime (i.e., {\it  dispersive Rydberg-EIT}), in which a weak nonlinear probe laser field couples the atomic ground state and an intermediate state (which has a large detuning), and a strong control laser field couples the intermediate state and a Rydberg state. Rydberg dark-state polariton, i.e., a coherent superposition of light field and atomic spin wave in the Rydberg gas, can form through Rydberg-EIT. We design an attractive potential (called Rydberg defect potential or Rydberg defect) by using gate photons stored in an atomic array occupying in another Rydberg state. We show that in such system the nonlocal nonlinear optical response of the system is largely modified, and the system may not only support nonlocal weak-light solitons, but also display interesting anomalous behaviors for QR when the optical solitons are scattered by the Rydberg defect. For simplicity, we shall call such analogue quantum reflection as quantum reflection in the following.

In Sec.~II, we present our physical model and derive a nonlinear envelope equation for the propagation of probe field under the condition of  Rydberg-EIT by employing an approach beyond mean-field approximation for many-atom correlations~\cite{Bai,ZhangQi,Hang,Bai2019}. This nonlinear envelope equation includes a term representing the local Rydberg defect potential contributed by the gate photons, and a term representing a nonlocal nonlinear attractive potential contributed by the Rydberg-Rydberg interaction. In Sec.~III, we demonstrate that the nonlinear envelope equation allows nonlocal optical soliton solutions, which have very low light power and may experience a sharp transition between reflection and transmission when scattered by the Rydberg defect. We find that the QR of the nonlocal optical solitons depend significantly on the incident velocity and incident power; in particular, the QR exhibits counter-intuitive (anomalous) behaviors due to Rydberg blockade effect, and is sensitive to the change of the nonlocality degree of the Kerr nonlinearity, which are very different from conventional QRs reported before. Thus, the QR in such system can be manipulated and controlled by actively adjusting system parameters. Additionally, thanks to the nonlocality of the Kerr nonlinearity, the system supports stable (2+1)-dimensional [(2+1)D] nonlocal optical solitons, which can also display a QR when scattered from the Rydberg defect.

The nonlocal optical solitons in the Rydberg atomic gas have advantages for  detailed investigation on QR. The reason is that such solitons are robust during propagation even in high spatial dimensions, and allow precise controls of the incident velocity, power, and nonlocality degree; furthermore, they have clean reflection and transmission when scattered from linear attractive potentials, which are observable under current experimental conditions. The results reported here open a new avenue for the study of Rydberg nonlinear optics, especially for the active control of QR in nonlocal nonlinear systems and for the deep exploration of the intriguing physical properties of Rydberg blockade and Rydberg defects, which have potential practical applications in optical information processing and transmission, including the design of novel optical devices (such as optical splitters, switchers, and transistors, etc.) that can work at weak-light level.

\section{Model}\label{model}

\subsection{Model and nonlinear envelope equation}\label{secIIa}

\begin{figure}
\centering
\includegraphics[width=0.5\textwidth]{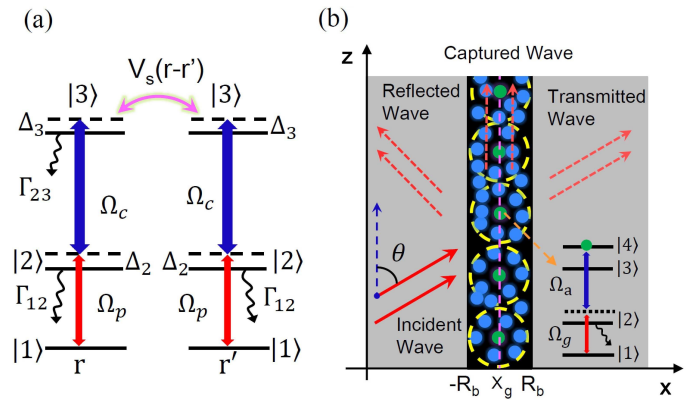}\\
\caption{\footnotesize(Color online) (a)~Ladder-type excitation scheme of the Rydberg-EIT, where a weak probe field couples the ground state $|1\rangle$ and the intermediate state $|2\rangle$ (with half Rabi frequency $\Omega_p$), and a strong control field couples $|2\rangle$ and the Rydberg state $|3\rangle$ (with half Rabi frequency $\Omega_c$), respectively. $\Delta_2$ and $\Delta_3$ are respectively one- and two-photon detunings, and $\Gamma_{12}$ and $\Gamma_{23}$ are respectively the spontaneous emission decay rates from $|2\rangle$ to $|1\rangle$ and $|3\rangle$ to $|2\rangle$. $V_s({\bf r}-{\bf r}^{\prime})$ is the Rydberg-Rydberg interaction potential between atoms at ${\mathbf r}$ and ${\mathbf r}^{\prime}$, respectively.
(b)~Schematic of the geometry for detecting the optical analogue of quantum reflection. The linear attractive potential (Rydberg defect) is prepared (via the use of another Rydberg-EIT) by the gate photons stored in atomic arrays in another Rydberg state $|4\rangle$ through a gate laser field (with half Rabi frequency $\Omega_g$) and an assistant laser field (with half Rabi frequency $\Omega_a$); see the level diagram shown by the inset. The region where the defect locates is illustrated by the domain with black color, where some Rydberg-blockade spheres of radius $R_b$ (with the solid circles representing atoms) are shown; the center of the defect is at position $x=x_g$ and $y=y_g$ along $z$ direction. The incident probe beam (with incident angle $\theta$) undergoes reflection and transmission, or captured when it collides with the defect.
}\label{fig1}
\end{figure}

We consider a cold, lifetime-broadened three-level atomic gas with a ladder-type EIT configuration~\cite{pritchard2}, as illustrated in Fig.~\ref{fig1}(a).
The electric field of the laser fields interacting with the atomic gas reads ${\bf E}({\bf r}, t)={\bf E}_p+{\bf E}_c=
\sum_{l=p,c}{\bf e}_l\, \mathcal{E}_l \,\exp [i({\bf k}_l
\cdot {\bf r}-\omega_l t)]+{\rm c.c.}$, where ${\bf e}_l$ $({\bf k}_l)$ is the unit polarization vector (wavevector) of the electric-field component with envelope $\mathcal{E}_l\,\,(l=p,c)$.
Here, a weak, spatially focused probe laser field ${\bf E}_p$ (with wave number $k_p=\omega_p/c$, angular frequency $\omega_p$, and half Rabi frequency $\Omega_p$) couples to the transition between the ground state $|1\rangle$ and the intermediate state $|2\rangle$; a strong, continuous-wave  control laser field ${\bf E}_c$ (with wave number $k_c=\omega_c/c$, angular frequency $\omega_c$, and half Rabi frequency $\Omega_c$) couples to the transition between intermediate state $|2\rangle$ and the Rydberg $|3\rangle$. $\Delta_2$ and $\Delta_3$ are respectively one- and two-photon detunings; $\Gamma_{12}$ and $\Gamma_{23}$ are respectively the spontaneous emission decay rates from $|2\rangle$ to $|1\rangle$ and $|3\rangle$ to $|2\rangle$. $V_s({\bf r}-{\bf r}^{\prime})\equiv -C_6^s/|{\mathbf r}-{\mathbf r}^{\prime}|^6$ is the Rydberg-Rydberg interaction potential between the atom at position ${\mathbf r}$ and the atom at position ${\mathbf r}^{\prime}$. The reasons for exploiting the Rydberg-EIT are to take the advantages of both the EIT and the Rydberg state. The former (EIT) can be used to suppress spontaneous emission from the short-lived intermediate state
$|2\rangle$, and the latter (Rydberg state) is long-lived and can be used to provide strong long-range Rydberg-Rydberg interaction and hence giant nonlocal Kerr nonlinearity for the probe field.

In order to investigate the QR of the probe field, a linear optical potential (called defect potential or defect) must be prepared initially, which can be realized by using the following method. We assume that some gate photons are stored (via a Rydberg-EIT) in the atomic array in another Rydberg state $|4\rangle$~\cite{Novikova3,Baur,Gorn1,WeibinLi,Mur1,Mur2} by using a gate laser field [with half Rabi frequency $\Omega_g$ that couples the levels $|1\rangle$ and $|2\rangle$; see the inset on the right hand side of Fig.~\ref{fig1}(b)] and an assistant laser field (with half Rabi frequency $\Omega_a$ that couples the levels $|2\rangle$ and $|4\rangle$). Note that such a scheme for preparing gate photons have been widely employed for realizing all-optical switchers and transistors with Rydberg atoms in Refs.~\cite{Baur,Gorn1,WeibinLi,Mur1,Mur2}, here we use them to produce the attractive defect potential. In this way, a Rydberg defect potential (i.e. the defect formed by the gate photons stored in the Rydberg state $|4\rangle$) for the probe field is created, by which the probe field will be scattered when it is incident on the defect. In Fig.~\ref{fig1}(b), the region of the defect is illustrated by the domain with black color, where some Rydberg-blockade spheres of radius $R_b$ (with the solid circles representing atoms) are shown; the center of the defect is at position ${\bf r}_g=(x_g, y_g, z)$, where $x_g$ and $y_g$ are fixed, but $z$ is arbitrary. The incident probe field (with incident angle $\theta$) will undergo reflection and transmission, and may be captured when scattered by the defect.

The dynamics of the system in interaction picture is described by the Hamiltonian $\hat{H}={\cal N}_{a}\int {\rm d}^3 r \,\hat{\cal H}(\mathbf{r},t)$, with $\hat{{\cal H}}(\mathbf{r},t)$ the Hamiltonian density and ${\cal N}_{a}$ the atomic density. Under electric-dipole and rotating-wave approximations, the Hamiltonian density reads
\begin{subequations}\label{Hami}
\begin{eqnarray}
&& \hat{{\cal H}}(\mathbf{r},t)= -\sum_{\alpha=1}^{3}\hbar\Delta_{\alpha}\hat{S}_{\alpha\alpha}(\mathbf{r},t)-\hbar\left[{\Omega_p\hat{S}_{12}(\mathbf{r},t)
+\Omega_c\hat{S}_{23}(\mathbf{r},t)}\right. \nonumber\\
&&\hspace{1.5cm}\left.{+{\rm H.c.}}\right]+\hat{{\cal H}}_{s}+\hat{{\cal H}}_{g}, \label{Hami1}\\
&& \hat{{\cal H}}_{s}(\mathbf{r},t)={\cal N}_{a}\int{\rm d}^3 r^{\prime}\hat{S}_{33}(\mathbf{r}^\prime,t)\hbar V_s(\mathbf{r}-\mathbf{r}^\prime)\hat{S}_{33}
(\mathbf{r},t),\label{Hami2}\\
&& \hat{{\cal H}}_{g}(\mathbf{r},t)={\cal N}_{g}\int{\rm d}^3{r}_g^{\prime}\hat{S}_{44}(\mathbf{r}_g^{\prime},t)\hbar V_d(\mathbf{r}-\mathbf{r}_g^{\prime})\hat{S}_{33}
(\mathbf{r},t), \label{Hami3}
\end{eqnarray}
\end{subequations}
where ${\rm d}^3 r =dx dy dz$, $\hat{S}_{\alpha\beta}=|\beta\rangle\langle\alpha|{\rm exp}i[({\bf k}_\beta-{\bf k}_\alpha)\cdot{\bf r}-(\omega_\beta-\omega_\alpha+\Delta_\beta-\Delta_\alpha)t]$ are transition operators related to the states $|\alpha\rangle$ and $|\beta\rangle$ $(\alpha,\beta = 1,2,3,4)$, satisfying the commutation relation
$\left[\hat{S}_{\alpha\beta}(\mathbf{r},t),
\hat{S}_{\mu\nu}(\mathbf{r}^{\prime},t)
\right]
=(1/{\cal N}_a)\delta(\mathbf{r}-\mathbf{r}^{\prime})
\left(\delta_{\alpha\nu}\hat{S}_{\mu \beta}(\mathbf{r},t)-\delta_{\mu \beta}\hat{S}_{\alpha\nu}(\mathbf{r}^{\prime},t)\right)$,
with $\hbar\omega_\alpha$ the eigenenergy of the level $|\alpha\rangle$, ${\cal N}_{g}$ the density of gate atoms, $\Delta_2=\omega_p-(\omega_2-\omega_1)$ and $\Delta_3=\omega_p+\omega_c-(\omega_3-\omega_1)$ being respectively the one-photon and two-photon detunings; $\Omega_p=(\mathbf{e}_p\cdot \mathbf{p}_{21})\mathcal{E}_p/(2\hbar)$ and $\Omega_c=(\mathbf{e}_c\cdot \mathbf{p}_{32})\mathcal{E}_c/(2\hbar)$ are respectively the half Rabi frequencies of the probe and control fields (with $\mathbf{p}_{\alpha\beta}$ the electric dipole matrix element associated with the transition from $|\beta\rangle$ to $|\alpha\rangle$); the Hamiltonian $\hat{{\cal H}}_{s}$ is the contribution due to Rydberg-Rydberg interaction, with $V_s({\bf r}-{\bf r}')=-C_6^s/|{\mathbf r}-{\bf r}^\prime|^6$ the van der Waals (vdW) interaction potential between the atom at ${\bf r}$ and the atom at ${\bf r}^\prime$ (for the atoms at ${\bf r}^\prime$ being in the Rydberg state $|3\rangle$; $C_6^s$ the dispersion coefficient); $\hat{{\cal H}}_{g}$ is the Hamiltonian describing the Rydberg-Rydberg interaction between the atom in the Rydberg state $|3\rangle$ and the atom in the Rydberg state $|4\rangle$ where the gate photons are stored, and hence the vdW interaction potential is $V_d({\bf r}-{\bf r}_g^{\prime})=-C_6^d/|{\mathbf r}-{\bf r}_g^{\prime}|^6$ the vdW interaction potential (for the atoms at ${\bf r}^\prime={\bf r}_g^\prime$ being in the Rydberg state $|4\rangle$; $C_6^d$ is the corresponding dispersion coefficient), with ${\bf r}_g^{\prime}=(x_g, y_g, z')$, where $x_g$ and $y_g$ are fixed and $z^{\prime}$ is arbitrary.

Based on the Hamiltonian $\hat{H}$ given above, we obtain the optical Bloch equation of one-atom density-matrix elements $\rho_{\alpha\beta}({\bf r},t)\equiv\langle {\hat S}_{\alpha\beta}({\bf r},t)\rangle$ with the form
\begin{subequations} \label{Eq21}
\begin{eqnarray}
&& i\frac{\partial }{\partial t}\rho_{11}-i\Gamma_{12}\rho_{22}-\Omega
_{p}\rho_{12}+\Omega _{p}^{\ast}\rho_{21}=0,\label{eq21} \\
&& i\frac{\partial }{\partial t}\rho_{22}-i\Gamma_{23}\rho
_{33}+i\Gamma_{12}\rho_{22}+\Omega _{p}\rho_{12}-\Omega
_{p}^{\ast}\rho_{21}\\\nonumber
&& \hspace{1cm}-\Omega_{c}\rho_{23}+\Omega_{c}^{\ast
}\rho_{32}=0,\label{eq22}\\
&& i\frac{\partial }{\partial t}\rho_{33}+i\Gamma_{23}\rho
_{33}+\Omega_{c}\rho_{23}-\Omega_{c}^{\ast}\rho_{32}=0,\label{eq23}
\end{eqnarray}
\end{subequations}
for diagonal elements, and
\begin{subequations} \label{Eq22}
\begin{eqnarray}
&& \left(i\frac{\partial }{\partial t}+d_{21}\right)
\rho_{21}-\Omega_{p}(\rho_{22}-\rho_{11})+\Omega_{c}^{\ast
}\rho_{31}=0,\label{eq24}\\
&&\left[i\frac{\partial }{\partial t}+d_{31}-\Delta_d(x,y)\right]\rho_{31}-\Omega
_{p}\rho_{32}+\Omega_{c}\rho_{21}
\nonumber\\
&& \hspace{1cm}-{\cal N}_a\int {\rm d}^3 r^{\prime} V_s(\mathbf{r}^\prime-\mathbf{r})\rho_{33,31}(\mathbf{r}^\prime, \mathbf{r},t)=0,\label{eq25}\\
&&\left[i\frac{\partial }{\partial t}+d_{32}-\Delta_d(x,y)\right] \rho_{32}-\Omega_{p}^{\ast}\rho_{31}-\Omega_{c}(\rho_{33}-\rho_{22})
\nonumber\\
&& \hspace{1cm} -{\cal N}_a\int {\rm d}^3 r^{\prime} V_s(\mathbf{r}^\prime-\mathbf{r})\rho_{33,32}(\mathbf{r}^\prime,\mathbf{r},t)=0,
\label{eq26}
\end{eqnarray}
\end{subequations}
for non-diagonal elements, where $d_{\alpha\beta}=\Delta_{\alpha}-\Delta_{\beta}+i\gamma_{\alpha\beta}$
($\alpha\neq \beta)$, with $\gamma_{\alpha\beta}\equiv(\Gamma_\alpha+\Gamma_\beta)/2
+\gamma_{\alpha\beta}^{\rm col}$. Here $\Gamma_\beta\equiv\sum_{\alpha<\beta} \Gamma_{\alpha\beta}$ with $\Gamma_{\alpha\beta}$ the spontaneous emission decay rate, and $\gamma_{\alpha\beta}^{\rm col}$  the dephasing rate between $|\alpha\rangle$ and $|\beta\rangle$. In Eqs.~(\ref{eq25}) and (\ref{eq26}), we have used the notation $\rho_{\alpha\beta,\mu\nu}(\mathbf{r}^\prime,\mathbf{r},t)\equiv\langle \hat{S}_{\alpha\beta}
(\mathbf{r}^\prime,t)\hat{S}_{\mu\nu}(\mathbf{r},t)\rangle$ for
two-atom density-matrix elements (i.e. two-atom correlators), whose dynamics is described by additional equations, which are related to the correlators of three-atom, four-atom, etc.~\cite{Bai,ZhangQi,Hang,Bai2019}. Explicit expressions of the equations for these many-atom correlators are lengthy and are omitted here for saving space.

The position-dependent detuning in Eqs.~(\ref{eq25}) and (\ref{eq26}) reads
\begin{eqnarray}\label{Delta}
\Delta_d(x,y)
&& ={-\cal N }_{gL}\int dz^\prime \frac{C_6^d}{[(x-x_g)^2+(y-y_g)^2+(z-z^\prime)^2]^3}\nonumber\\
&& =-\frac{3\pi {\cal N}_{gL} C_6^d}{8(|x-x_g|^5+|y-y_g|^5)},
\end{eqnarray}
with ${\cal N}_{gL}$ the linear density of the atoms at the state $|4\rangle$. $\Delta_d(x,y)$ is contributed by the atoms at the Rydberg state $|4\rangle$ where the gate photons are stored,
which will play the role of the Rydberg defect potential for the scattering of the probe field, see below.

For investigating the scattering of the probe field by the Rydberg defect potential, we assume the size of the atomic gas is much larger than the Rydberg blockade radius $R_b\,(\equiv [|C_6^d d_{21}|/(2|\Omega_c|^2)]^{1/6})$. The propagation of the probe field is described by the Maxwell equation, which under slowly varying amplitude approximation is reduced into
\begin{equation}\label{Max}
i\left(\frac{\partial}{\partial z}+\frac{1}{c}\frac{\partial}{\partial t} \right) \Omega_{p}+\frac{1}{2k_p}\left(\frac{\partial^2}{\partial x^2}+\frac{\partial^2}{\partial y^2}\right) \Omega_p+\frac{k_p}{2}\chi_p\Omega_p=0,
\end{equation}
where $\chi_p={\cal N}_a ( {\bf e}_p\cdot {\bf p}_{12})^2 \rho_{21}/(\varepsilon_0 \hbar\Omega_{p})$ is the optical susceptibility, with $\rho_{21}({\bf r},t)\equiv\langle {\hat S}_{21}({\bf r},t)\rangle$ the coherence between the states $|1\rangle$ and $|2\rangle$.
For simplicity, we assume that the probe field has a long time duration, so that the system works in a steady state, and hence the time derivatives in the Maxwell-Bloch (MB) Eqs.~(\ref{Eq21}), (\ref{Eq22}), and (\ref{Max}) are negligible.

For a relatively weak probe field, the population in atomic levels changes not much when the probe field is applied to the system, and hence a perturbation expansion beyond mean-field approximation for many-atom correlations can be employed to solve the Bloch equation Eqs.~(\ref{Eq21}) and (\ref{Eq22})~\cite{Bai,ZhangQi,Hang,Bai2019}.
The expression of the nonlinear optical susceptibility of the probe field exact to the third order of the perturbation expansion is given by [see Eq.~(\ref{chiapp}) of Appendix~\ref{ap1}]
\begin{equation}\label{chi}
\chi_p\simeq\chi_p^{(1)}+\chi_{p,1}^{(3)}|\mathcal{E}_p|^2 +\int {\rm d}^3 r^{\prime}\chi_{p,2}^{(3)}(\mathbf{r}^\prime-\mathbf{r})
|\mathcal{E}_p(\mathbf{r}^\prime)|^2,
\end{equation}
with $\chi_p^{(1)}={\cal N}_a|\mathbf{p}_{12}|^2a_{21}^{(1)}/(\varepsilon_0\hbar)$, $\chi_{p,1}^{(3)}={\cal N}_a|\mathbf{p}_{12}|^4 a_{21,1}^{(3)}/(\varepsilon_0\hbar^3)$, and $\chi_{p,2}^{(3)}={\cal N}_a^2|\mathbf{p}_{12}|^4a_{21,2}^{(3)}/(\varepsilon_0\hbar^3)$.
Explicit expressions of $a_{21}^{(1)}$, $a_{21,1}^{(3)}$, and $a_{21,2}^{(3)}$
are given by Eqs.~(\ref{A211}), (\ref{A2113}), and (\ref{A2123}) of the Appendix~\ref{ap1}, respectively.

For simplicity, we assume that the spatial length of the probe beam in $z$ direction is much larger than the range of Rydberg-Rydberg interaction, so that a local approximation along the $z$ direction can be made~\cite{Sev}. Hence, the last term of the susceptibility  $\int {\rm d}^3 r^{\prime}\chi_{p,2}^{(3)}(\mathbf{r}^\prime-\mathbf{r})
|\Omega_p(\mathbf{r}^\prime)|^2\simeq \int{dx^\prime dy^\prime\tilde{\chi}_{p,2}^{(3)}(x-x^\prime,y-y^\prime)
|\Omega_p(x^\prime,y^\prime,z)|^2}$, with $\tilde{\chi}_{p,2}^{(3)}(x-x^\prime,y-y^\prime)
=\int{dz^\prime\chi_{p,2}^{(3)}(\mathbf{r}^\prime-\mathbf{r})}$.
Then the Maxwell Eq.~(\ref{Max}) is reduced into
\begin{eqnarray}\label{Max2}
&& i\frac{\partial \Omega_{p}}{\partial z}+\frac{1}{2k_p}\nabla_{\perp}^2 \Omega_p+\frac{k_p}{2}\chi_p^{(1)}\Omega_p+\frac{bk_p}{2}
\left({\chi_{p,1}^{(3)}|\Omega_p|^2} \right.\\
&& \left.{+\int{dx^\prime dy^\prime\tilde{\chi}_{p,2}^{(3)}(x-x^\prime,y-y^\prime)
|\Omega_p(x^\prime,y^\prime,z)|^2}}\right)\Omega_p=0,\nonumber
\end{eqnarray}
where $\nabla_{\perp}^2=\partial^2/\partial x^2+\partial^2/\partial y^2$,
$b=(\hbar/|\mathbf{p}_{12}|)^2$ is coupling coefficient, $\chi_p^{(1)}$ is linear susceptibility [proportional to the position-dependent detuning $\Delta_d(x,y)$ given by Eq.~(\ref{Delta}) and related to the Rydberg defect potential], $\chi_{p,1}^{(3)}$ is local third-order nonlinear susceptibility, and $\tilde{\chi}_{p,2}^{(3)}$ is the kernel of nonlocal third-order nonlinear susceptibility (contributed by the long-range Rydberg-Rydberg interaction).
Note that the local nonlinear susceptibility is proportional to the atomic density  (i.e. $\chi_{p,1}^{(3)}\propto{\cal N}_a$), and it vanishes when the two-photon detuning $\Delta_3=0$; however, the nonlocal nonlinear susceptibility has a nonlinear dependence on the  atomic density (i.e. $\chi_{p,2}^{(3)}\propto{\cal N}_a^2$), and it is non-zero for $\Delta_3=0$. Thus one sees that the nonlocal Kerr nonlinearity can be much greater than the local one for a large atom density.

For the convenience of later calculations, we convert Eq.~(\ref{Max2}) into the dimensionless form
\begin{eqnarray}\label{nonlinearEq}
&&i\frac{\partial U}{\partial s}=-\left( \frac{\partial^2 }{\partial \xi^2}+\frac{\partial^2 }{\partial \eta^2}\right)U+V(\xi,\eta)U+\left[{W_1|U|^2}
\right.\nonumber\\
&&\hspace{0.7cm}\left.{+\int{d\xi^\prime d\eta^\prime W_2(\xi-\xi^\prime,\eta-\eta^\prime)|U(\xi^\prime, \eta^\prime, s)|^2}}\right]U,
\end{eqnarray}
where $U=\Omega_p/\Omega_{p0}$, $(\xi, \eta)=(x, y)/R_0$, and $s=z/(2L_{\rm diff})$  are dimensionless variables, with $\Omega_{p0}$, $R_0$, and $L_{\rm diff}\equiv k_p R_0^2$ representing respectively typical half Rabi frequency, width of the probe beam, and diffraction length of the probe field; $V\equiv-k_p^2 R_0^2\,\chi_p^{(1)}(x,y)$, $W_1\equiv -bk_p^2 R_0^2U_0^2\,\chi_{p,1}^{(3)}$, and
$W_2\equiv-bk_p^2 R_0^4U_0^2\,\tilde{\chi}_{p,2}^{(3)}$ are respectively
dimensionless linear potential (i.e. the Rydberg defect potential contributed by the gate photons stored in the Rydberg state $|4\rangle$), coefficient of local Kerr nonlinearity  (contributed by the atom-photon interaction), and coefficient of nonlocal
Kerr nonlinearity (contributed by the Rydberg-Rydberg interaction).
The term in the square bracket on the right hand side of Eq.~(\ref{nonlinearEq}) can be regarded as a nonlinear potential for the propagation of the probe field. Due to the strong Rydberg-Rydberg interaction, the nonlocal Kerr nonlinearity is much larger than the local one, and thus the term  $W_1|U|^2 U$ is negligible (see Sec.~\ref{secIIIa}).

\subsection{Attractive Rydberg defect potential}\label{secIIb}

We first discuss the physical properties of the Rydberg defect potential $V(\xi,\eta)$ in Eq.~(\ref{nonlinearEq}). To be concrete, we take laser-cooled strontium ($^{88}$Sr) atomic gas as a realistic candidate for our theoretical model described above. The energy-levels shown in Fig.~1(a) are selected to be $|1\rangle=|5s^2\,^1S_0\rangle$, $|2\rangle=|5s5p ^1P_1\rangle$, $|3\rangle=|5sns^1S_0\rangle$, and $|4\rangle=|5sn^\prime s^1S_0\rangle$. For the principal quantum number $n=n^\prime=60$, the dispersion coefficients are $C_6^s=C_6^d\simeq 2\pi\times 81.6\,{\rm GHz} \mu {\rm m}^6$~\cite{Mauger}. For such choice, the vdW interaction is isotropically attractive, and hence the Rydberg defect potential $V(\xi,\eta)$  is attractive and the nonlocal Kerr nonlinearity is a self-focusing one. The spontaneous emission rates read $\Gamma_{12}=2\pi\times16\,\,{\rm MHz}$ and $\Gamma_{23}=2\pi\times16.7\,{\rm kHz}$. Other system parameters are given by ${\cal N}_a=3\times10^{10}$\,cm$^{-3}$, ${\cal N}_{gL}=200$\,cm$^{-1}$, $\Delta_3=0$, $R_0=10$\,$\mu{\rm m}$, and $\Omega_c=2\pi\times16\,{\rm MHz}$; especially, a large
one-photon detuning $\Delta_2=-2\pi\times160\,{\rm MHz}$ is taken
to make the system work in a regime of dispersive Rydberg-EIT.

\begin{figure}
\centering
\includegraphics[width=0.5\textwidth]{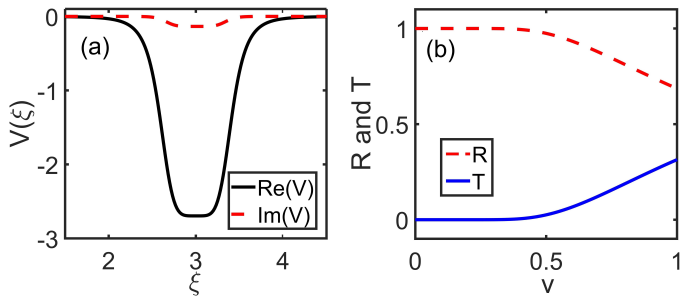}\\
\caption{\footnotesize(Color online) (a)~Dimensionless attractive Rydberg defect potential $V$ in (1+1)D as a function of dimensionless coordinate $\xi=x/R_0$ for $\Delta_2=-2\pi\times160\,{\rm MHz}$. The solid black line and the dashed red line are the real part of Re$(V)$ and the imaginary part Im$(V)$, respectively.
(b)~Reflection coefficient $R$ (dashed red line) and transmission coefficient $T$ (solid blue line) of a linear probe field as functions of dimensionless incident velocity $v$.
}\label{fig2}
\end{figure}

For simplicity, we consider (1+1)D Rydberg defect potential $V=V(\xi)$ by taking $x_g=30\,\mu{\rm m}$  and $y_g$ arbitrary (a generalization to (2+1)D case will be discussed Sec.~\ref{secIIId}).
Based on the above parameters, one obtains the blockade radius $R_b\simeq8\mu{\rm m}$ and the diffraction length $L_{\rm diff}\simeq1.36$\,mm. Fig.~\ref{fig2}(a)
shows $V$ as a function of $\xi=x/R_0$. The real part [Re$(V)$] and imaginary parts [Im$(V)$] are plotted for $\Delta_2\gg\Gamma_{12}$ by the solid black line and the dashed red line, respectively. We see that Im$(V)$ is much smaller than Re$(V)$, which means that the optical absorption of the probe field is negligible, originated by the EIT effect and the condition of large one-photon detuning; moreover, Re$(V)$ is an attractive potential well and there is a saturation near $\xi=x_g/R_0=3$, which is due to the Rydberg blockade effect that suppresses the excitation of atoms to the Rydberg state and hence causes the potential $V$ to saturate to a finite value.

The (1+1)D reflection and transmission of the probe field can be studied using Eq.~(\ref{nonlinearEq}) by taking $\partial/\partial\eta=0$, $V=V(\xi)$, and $W_1=W_2=0$. Assuming the probe beam is incident to the Rydberg defect potential $V(\xi)$ from left hand side [see Fig.~\ref{fig1}(b)]. Generally, full reflection, transmission, or trapping will occur. These scattering behaviors can be respectively described by the reflection coefficient $R$, transmission coefficient $T$, and trapping coefficient $L$, defined by~\cite{Khawaja}
\begin{eqnarray}\label{TLR}
&& R=\frac{\int^{\xi_l}_{-\infty} d\xi |U(\xi,z=L_{\rm m})|^2}{\int^{+\infty}_{-\infty} d\xi |U(\xi,z=0)|^2},\,
T=\frac{\int^{+\infty}_{\xi_r} d\xi |U(\xi, z=L_{\rm m})|^2}{\int^{+\infty}_{-\infty} d\xi |U(\xi,z=0)|^2}, \nonumber\\
&& L=\frac{\int^{\xi_r}_{\xi_l} d\xi |U(\xi,z=L_{\rm m})|^2}{\int^{+\infty}_{-\infty} d\xi |U(\xi,z=0)|^2}, \,
\end{eqnarray}
with $L_{\rm m}$ being the length of the medium along the $z$ direction, $\xi_l$ ($\xi_r$) being the position on the $\xi$ axis at which the influence of the potential on the left (right) hand side of the defect potential is negligible, with $T+L+R=1$. Fig.~\ref{fig2}(b) shows the result of a numerical simulation on the (1+1)D reflection and transmission of a linear probe field, with the dashed red (solid black) line denoting the reflection (transmission) coefficient as a function of the dimensionless incident velocity of the probe-field photons $v$~\cite{note3}. One sees that as $v$ is increased a QR occurs for the incident probe beam and there exists a smooth transition between the reflection and the transmission.
Note that in the present study we are interested only in the QR of nonlocal solitons, the linear probe field is not trapped when it is incident upon the Rydberg defect potential with the system parameters considered here (i.e. trapping coefficient $L$ is zero for linear probe fields).

\section{QR of nonlocal optical solitons}\label{secIII}

\subsection{(1+1)D nonlocal optical solitons}\label{secIIIa}

We now turn to consider the nonlinear propagation of the probe field. Based on the system parameters given in the beginning of Sec.~\ref{secIIb}, together with $\Omega_{p0}=33$\,MHz, we can estimate the nonlinear coefficients $W_1$ and $W_2$ in the nonlinear envelope equation (\ref{nonlinearEq}), which are respectively given by $W_1\approx(4.95+i0.49)\times10^{-10}$ and $\int{d\xi d\eta W_2(\xi,\eta)}\approx-3.17-i0.08$. We see that due to the strong Rydberg-Rydberg interaction the nonlocal Kerr nonlinearity is ten orders of magnitude greater than the local one and hence the later can be neglected safely.

\begin{figure}
\centering
\includegraphics[width=0.5\textwidth]{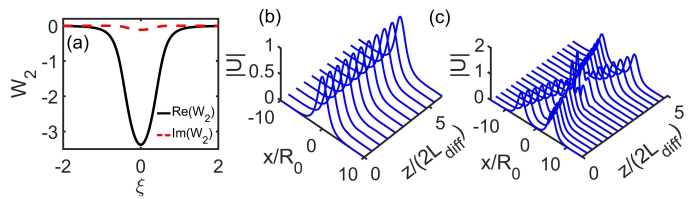}\\
\caption{\footnotesize(Color online)
(a)~Spatial distribution of $W_2$ (i.e. the dimensionless coefficient of the nonlocal Kerr nonlinearity) as a function of $\xi=x/R_0$, in which the solid black (dashed red) line is its real part Re$(W_2)$ [imaginary part Im$(W_2)$] for $\eta=0$.
(b)~Nonlocal optical soliton, by taking the its dimensionless amplitude $|U|$ as a function of $x/R_0$ and $z/(2L_{\rm diff})$.
(c)~Collision between two nonlocal optical solitons.}
\label{soliton}
\end{figure}

Fig.~\ref{soliton}(a)
illustrates the spatial distribution of $W_2(\xi,\eta)$ as a function of $\xi$, with the solid black (dashed red) line representing its real part Re$(W_2)$ [imaginary part Im$(W_2)$] for $\eta=0$. We see that Im$(W_2)$ is much smaller than Re$(W_2)$, which is also due to the EIT effect and the large one-photon detuning; additionally, Re$(W_2)$ is also an attractive potential well [and hence the nonlinear potential in Eq.~(\ref{nonlinearEq}) is an attractive one), which comes from our choice $^{88}$Sr atoms, for which the $n ^1S_0$ state is isotropically attractive ($C_{6}^s> 0$). Such choice allows us to obtain bright soliton solutions and investigate their QR property of the solitons.

We investigate first the propagation and scattering of (1+1)D nonlocal optical solitons, which can be obtained by the following assumptions: (i)~the probe field has a wide distribution in $y$ direction so that its $y$ dependence can be neglected, and (ii)~the gate photon distribution in the Rydberg state $|4\rangle$ is prepared to be independent of $y$. Hence in Eq.~(\ref{nonlinearEq}) one has $V=V(\xi)$, $W_2=W_2(\xi)$, and the term $\partial^2 U/\partial \eta^2$ can be neglected. As a result, after disregarding the negligible local nonlinear potential $W_1|U|^2$,  Eq.~(\ref{nonlinearEq}) is simplified into
\begin{equation}\label{nonEq1D}
i\frac{\partial U}{\partial s}=-\frac{\partial^2 U}{\partial \xi^2}+V(\xi)U+V_{\rm non}(\xi, U)U,
\end{equation}
with the nonlocal nonlinear potential given by $V_{\rm non}(\xi, U)=\int{d\xi^\prime W_2(\xi-\xi^\prime)|U(\xi^\prime)|^2}$.

In the absence of the linear attractive potential $V$,  Eq.~(\ref{nonEq1D}) allows various nonlocal optical soliton solutions. Plotted in Figure~\ref{soliton}(b) is the propagation of a nonlocal optical soliton by taking its amplitude $|U|$ as a function of $x/R_0$ and $z/(2L_{\rm diff})$, obtained by exploiting split-step Fourier method~\cite{Yang} with the initial condition $U(\xi, s=0)={\rm sech}(\xi)$. We see that the nonlocal optical soliton is robust during propagation. To test the stability of the soliton, a collision between two such solitons is also studied, with the result shown in Fig.~\ref{soliton}(c). The initial condition for the collision calculation is given by $U(\xi, s=0)={\rm sech}(\xi+5){\rm exp}(i\xi)+{\rm sech}(\xi-5){\rm exp}(-i\xi)$. One sees that the both solitons resume their original shapes after the collision.

The peak power $P_{\rm max}$ for generating the nonlocal optical soliton can be estimated using Poyntings vector~\cite{Hang,Bai2019}, which, based on the system parameters give above, reads
\begin{equation}
P_{\rm max}\simeq1.5\,\, {\rm nW},
\end{equation}
with corresponding average peak intensity given by $I_{\rm max}\simeq1.2$ mW cm$^{-2}$. Consequently, for generating such nonlocal optical soliton only a very weak light power is needed. This is in contrast with cases of nonresonant media (such as optical fiber), where much higher light power is required for the formation of optical solitons.
We stress that the choice of $|n ^1S_0\rangle$ states of $^{88}$Sr atoms stated above is to obtain the self-focusing Kerr nonlinearity for balancing the diffraction effect of the system. Except for $^{88}$Sr, a recent study showed that the $|n ^3S_0\rangle$ states of $^{87}$Sr atoms can also provide attractive Rydberg-Rydberg interaction for some principal quantum numbers $n$~\cite{Robicheaux2019}, which provides another Rydberg gas that supports the formation of nonlocal bright solitons.

\subsection{QR of (1+1)D nonlocal optical solitons}\label{secIIIb}

We now study what will happen when the nonlocal optical soliton is scattered by the Rydberg defect potential $V(\xi)$. We assume that the soliton is incident from left side and has the form  $U(\xi, z=0)=A\,{\rm sech}[(\xi-\xi_0)/w_0]{\rm exp}(iv\xi)$, with $A=1$, $w_0=1$.
The initial position of the soliton, $\xi_0=x_0/R_0$, is chosen to be far from the Rydberg defect by taking $\xi_0=-2$.

\begin{figure}
\centering
\includegraphics[width=0.5\textwidth]{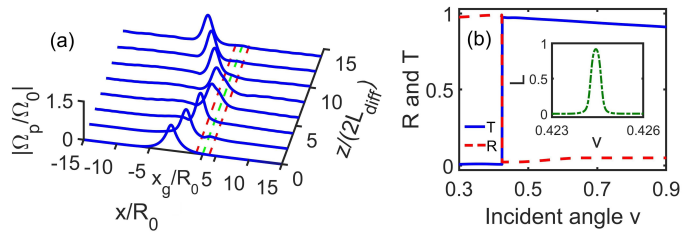}\\
\caption{\footnotesize(Color online) Quantum reflection of (1+1)D nonlocal optical soliton.
(a)~Soliton-amplitude function $|U|$ for $\Omega_{p0}=33$\,MHz  as a function of $x/R_0$ and $z/(2L_{\rm diff})$ when it is incident from the left hand side of the attractive Rydberg defect. The soliton is almost fully reflected by the defect with a small incident velocity $v=0.35$.
The dashed green line denotes the central position of the defect, and the width of the defect is marked by two dashed red lines.
(b) The sharp transition between nearly full reflection and nearly full transmission of the soliton, with reflection coefficient $R$  (dashed red line) and transmission coefficient $T$  (solid blue line) as functions of incident velocity $v$. The inset (dashed-dotted green curve) gives the trapping coefficient $L$ of the soliton as a function of $v$.
}\label{quantumR}
\end{figure}

Fig.~\ref{quantumR}(a) shows
the result of a numerical simulation for such scattering with a small incident velocity $v=0.35$, by taking for $\Omega_{p0}=33$\,MHz and the soliton-amplitude function $|U|$ as a function of $x/R_0$ and $z/(2L_{\rm diff})$. In the figure, the dashed green line denotes the central position of the defect, and the width of the defect (i.e. the Rydberg blockade region) is marked by two dashed red lines. We see that the nonlocal soliton is almost fully reflected, which is a typical character of QR since the potential $V$ is attractive. Physically, the QR can be understood as a specific interference phenomenon of incoming and outgoing waves~\cite{Miret,Petersen} when the nonlocal optical soliton interacts with the attractive Rydberg defect potential, absent for the scattering of classical particles because such phenomenon cannot be predicted  based on the theory of Newtonian mechanics.

To acquire a deep understanding on the QR of the nonlocal soliton, a further numerical simulations is carried out for different incident velocity $v$. Shown in Fig.~\ref{quantumR}(b) is the result of the transmission coefficient $T$ (solid blue line) and reflection coefficient $R$ (dashed red line) as functions of $v$ for $\Omega_{p0}=33$\,MHz. We see that, compared to the linear case obtained in Fig.~\ref{fig2}(b), the dependence of the reflection coefficient $R$ and the transmission coefficient $T$ on the incident velocity $v$ is drastically changed for the scattering of the nonlocal optical soliton. In particular, a pronounced new character appears due to the nonlocal nonlinear interaction, i.e. a {\it sharp} transition between the reflection and the transmission is observed with a well-defined critical velocity $v=v_c=0.42$. For $v<v_c$  the soliton scattering is dominated by nearly a full reflection; however, for $v>v_c$  the soliton scattering is dominated by nearly a full transmission. In the respective dominant regimes, the reflection (or transmission) of the soliton can be larger than $97\%$.
Note that the nonlocal soliton can experience a self-trapping near the critical velocity $v_c$. The inset of the panel (b) in the figure (dashed-dotted green line) gives the trapping coefficient $L$ of the soliton as a function of $v$.

The QR phenomenon of the nonlocal soliton described above
can be explained by using a two-mode picture~\cite{Goodman,Lee}. Since the attractive Rydberg defect potential $V(\xi)$ [Fig.~\ref{fig2}(a)] allows bound states, when the low-velocity soliton approaches and overlaps with the defect potential the solution $U$ of Eq.~(\ref{nonEq1D}) can be taken as a superposition of the soliton mode $U_S$ and the trapped (bound state) mode $U_T$, i.e. $U(\xi,s)=U_S(\xi,s)+U_T(\xi,s)$. Then off-diagonal terms, i.e. $U_S^\ast(\xi,s)V(\xi)U_T(\xi,s)$ and $U_T^\ast(\xi,s)V(\xi)U_S(\xi,s)$,  appear in the energy-density expression of the system, which will lead to a repulsive force between the two modes if they are out of phase, and hence a destructive interference arises so that a full reflection of the soliton occurs once the repulsive force due to this destructive interference overcomes the attractive force provided by the Rydberg defect potential. Note that the nonlocal nonlinear potential $V_{\rm non}(\xi, U)$ in the present system plays a specific role for the QR, which results in new, anomalous characters for the QR not found in conventional systems studied before, see the next subsection.

\subsection{Anomalous QR of the (1+1)D nonlocal optical solitons
and their active control}\label{secIIIc}

We now investigate what will happen for the nonlocal optical soliton scattering if the incident velocity of the soliton is fixed but its incident intensity $\Omega_{p0}$ is changed. For illustration, we take $v= 0.5$ and other system parameters the same as those used in Fig.~\ref{quantumR}.

\begin{figure}
\centering
\includegraphics[width=0.5\textwidth]{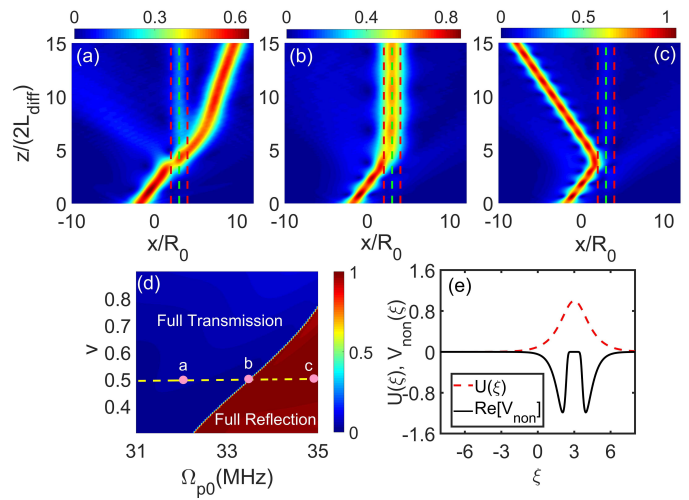}\\
\caption{\footnotesize(Color online) Anomalous scattering behaviors of the (1+1)D nonlocal optical solitons.
Panels (a), (b), and (c) show the nearly full transmission, trapping, and reflection of the soliton when the soliton is scattered by the Rydberg defect potential, with the dimensionless incident velocity fixed ($v= 0.5$) and the incident half Rabi frequency taken to be $\Omega_{p0}=32.3\,{\rm MHz}$, $\Omega_{p0}=33.4\,{\rm MHz}$, $\Omega_{p0}=35\,{\rm MHz}$, respectively.
As in Fig.~\ref{quantumR}(a), here the dashed green line denotes the central position of the defect, and the width of the defect is marked by two dashed red lines.
(d)~``Phase diagram'' of the soliton scattering, by taking the
reflection coefficient $R$ as a function of the incident velocity $v$ and  $\Omega_{p0}$. The domain with the blue (purple) color is the one of nearly full transmission (reflection). The line between the domains of the nearly full transmission and the nearly full reflection  is a crossover one, where soliton trapping occurs. Dots (with pink color) ``a'', ``b'', and ``c'' indicate the values of reflection coefficient $R$ for the cases shown in panels (a), (b) and (c), respectively.
(e)~Solid black line: nonlinear potential $V_{\rm non}(\xi, U)$ as a function of $\xi$; Dashed red line: soliton-amplitude function $U$; both $V_{\rm non}$ and $U$ are dimensionless.
}\label{fig5}
\end{figure}

Shown in panels (a), (b) and (c) of Fig.~\ref{fig5}
is the numerical result of the soliton scattering when it is incident from the left side of the Rydberg defect by choosing $\Omega_{p0}=32.3\,{\rm MHz}$, $33.4\,{\rm MHz}$, and $35\,{\rm MHz}$, respectively. From the figure, we see that the soliton gets a nearly full transmission for the weak incident power [$\Omega_{p0}=32.3\,{\rm MHz}$, panel (a)], and a full reflection for the strong incident power [$\Omega_{p0}=35\,{\rm MHz}$, panel (c)]. For the intermediate incident power [$\Omega_{p0}=33.4\,{\rm MHz}$, panel (b)], most of the incoming power of the soliton is captured by the defect with trapped probability $L=92\%$. It seems that such scattering behavior of the soliton is counter-intuitive (anomalous), since generally a full transmission (reflection) should occur for large (small) incident power.

To get a general picture, a further numerical simulation  is carried out for acquiring a ``phase diagram'' of the soliton scattering by taking the reflection coefficient $R$ as a function of $v$ and $\Omega_{p0}$, with the result of the simulation presented in Fig.~\ref{fig5}(d). In the figure, the domain with the blue color and the domain with purple color are regions for
nearly full transmission and nearly full reflection, respectively. The line between these two domains is the boundary representing the crossover from the nearly full transmission to the nearly full reflection, where soliton trapping occurs. Dots (with pink color) ``a'', ``b'', and ``c'' indicate the values of reflection coefficient $R$ for the cases shown respectively in panels (a), (b) and (c) of the figure. From the figure, we see that the full transmission (reflection) for small (large) $\Omega_{p0}$ is not a particular but general behavior in our system.

The anomalous behavior of the soliton scattering shown here can be explained as follows. Note that, due to the contribution of the stored gate photons in the Rydberg defect, the nonlinear potential $V_{\rm non}(\xi,U)$ in Eq.~(\ref{nonEq1D}) has a shape of ``double well'', shown in Fig.~\ref{fig5}(e) by the solid black line as a function of $\xi$ (the Rydberg defect is assumed to locate at $\xi\equiv x/R_0=3$). The soliton-amplitude function $U$ is also illustrated by the dashed red line~\cite{note4}. From the figure, we see that the nonlinear potential $V_{\rm non}(\xi,U)$ is attractive far from the Rydberg defect and repulsive close to the defect. We have thus the following conclusions:
(i)~Far from the Rydberg defect, the linear potential $V(\xi)$ is zero and the nonlinear potential $V_{\rm non}(\xi, U)$ is attractive. The probe beam can form a soliton [through solving Eq.~(\ref{nonEq1D})] by using  a suitable incident condition.
(ii)~Near the Rydberg defect, $V(\xi)$ is non-zero and attractive [i.e. a potential well; see Fig.~\ref{fig2}(a)]; however, the nonlinear potential $V_{\rm non}(\xi,U)$  is a potential barrier and hence it is repulsive~\cite{note5} [see Fig.~\ref{fig5}(e)]. If the incident probe-beam intensity is small (i.e. $\Omega_{p0}$ is small), the nonlinear potential $V_{\rm non}(\xi,U)$ plays no significant role and hence negligible. In this case, if the incident probe-beam velocity $v$ is not small [e.g. $v=0.5$ used in panels (a), (b), and (c) of Fig.~\ref{fig5}], the probe beam displays no full reflection and trapping but a full transmission, which is just the phenomenon observed in Fig.~\ref{fig5}(a). Nevertheless, if $\Omega_{p0}$ is large, the nonlinear potential $V_{\rm non}(\xi,U)$ plays a significant role and a full reflection occurs for the probe beam with the same incident velocity $v$. Based on such explanation, the ``phase diagram'' of the scattering of the probe beam given in Fig.~\ref{fig5}(d) can be well understood physically.

\begin{figure}
\centering
\includegraphics[width=0.5\textwidth]{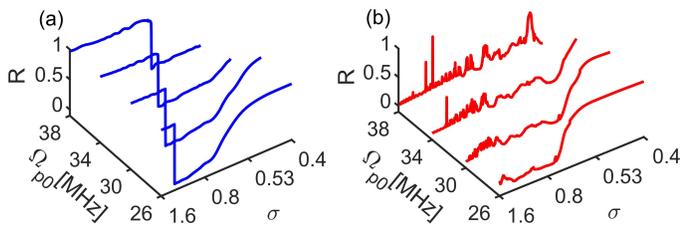}\\
\caption{\footnotesize(Color online) Active control of the QR of the nonlocal optical solitons.
(a)~Reflection coefficient $R$ as a function of $\Omega_{p0}$ and $\sigma$ (nonlocality degree of the Kerr nonlinearity).
(b)~Reflection $R$ as a function of $\Omega_{p0}$ and $\sigma$, for the case when the Rydberg defect is assumed to have no influence on the nonlocal Kerr nonlinearity. For detail, see the text.
}\label{fig6}
\end{figure}

Note that the nonlinear repulsive potential $V_{\rm non}(\xi, U)$ depends not only on the probe-field intensity but also on the nonlocality degree of the Kerr nonlinearity defined by~\cite{Bai2019}
\begin{equation}
\sigma=R_b/R_0.
\end{equation}
Shown in Fig.~\ref{fig6}(a)
is the result of a numerical simulation on the reflection coefficient $R$ of the nonlocal soliton by taking the Rydberg blockade radius $R_b=8\,\mu{\rm m}$ and varying the radius of the beam $R_0$ from $5\,\mu{\rm m}$ to $20\,\mu{\rm m}$.  It reveal that reflection $R$ is strongly dependent not only on the probe-field intensity (i.e. $\Omega_{p0}$), but also on the nonlocality degree of the Kerr nonlinearity (i.e. $\sigma$). Thus, in addition to $\Omega_{p0}$, the parameter $\sigma$ can be taken to manipulate and control the scattering of the nonlocal soliton. On the contrary, one can also employ the scattering data of the soliton to investigate the physical property of the system, including the measurement of the nonlocality degree of the Kerr nonlinearity (and hence the radius of the Rydberg blockade).

We stress that there is another new feature on the soliton scattering in our system, i.e., the existence of the Rydberg defect has a strong influence on the nonlocal Kerr nonlinearity, which makes the QR of the nonlocal optical soliton very different from those reported before. To see this clearly, a simulation is
carried out by taking $\Delta_d=0$ in $W_2$ [i.e. the coefficient describing the nonlocal Kerr nonlinearity; see Eq.~(\ref{nonlinearEq})], with the result given in Fig.~\ref{fig6}(b). We find that for weak probe field ($\Omega_{p0}=26$\,MHz), the reflection coefficient $R$ is the same as that in panel (a) (where $\Delta_d\neq 0$ in $W_2$) for
$0.4 < \sigma < 0.8$. However,  the strong reflection for $\sigma > 1.4$ shown in panel (a) is absent in panel (b), which means us that the influence of the Rydberg defect on the nonlinear potential $V_{\rm non}(\xi, U)$ is significant
for large $\sigma$, by which a strong repulsive force appears, acts on the nonlocal soliton, and results in a large reflection coefficient $R$. By comparing with panel (a) and panel (b) for different $\Omega_{p0}$, we see that, as a function of $\sigma$, the reflection coefficient $R$ has a minimum in panel (a) but the minimum is absent in  panel (b); generally speaking,
the soliton reflection in the case of panel (b) is much weaker than that in panel (a). These interesting anomalous behaviors of the soliton scattering and their controllability found here might be useful to design power and nonlocality dependent optical splitters, switches, and transistors.

\subsection{Scattering of (2+1)D nonlocal optical solitons}\label{secIIId}

In general, high-dimensional solitons are unstable during propagation~\cite{Malomed,Kartashov}. However, as demonstrated recently~\cite{Sev,Hang,Bai2019}, high-dimensional optical solitons are quite stable in Rydberg atomic gases due to the existence of the nonlocal Rydberg-Rydberg interaction. Here we show that (2+1)D nonlocal optical solitons
may have QR when they collide with a 2D Rydberg defect~\cite{note7} in the present system.

To this end, a numerical simulation on the scattering of a (2+1)D soliton from the attractive Rydberg defect is implemented based on Eq.~(\ref{nonlinearEq}), by taking the incident wavefunction $U(\xi,\eta,s=0)={\rm exp}[-(\xi^2+\eta^2)+i(v_{\xi}\xi+v_{\eta}\eta)]$, with $v_{\xi}$ and $v_{\eta}$ the dimensionless incident velocities of the soliton in $\xi$ and $\eta$ directions, respectively. As an example, in the simulation we have taken $v_{\xi}=v_{\eta}=0.5$, $R_0=10\,\mu$m, ${\cal N}_a=3\times10^{10}$\,cm$^{-3}$, and other parameters given previously.

\begin{figure}
\centering
\includegraphics[width=0.5\textwidth]{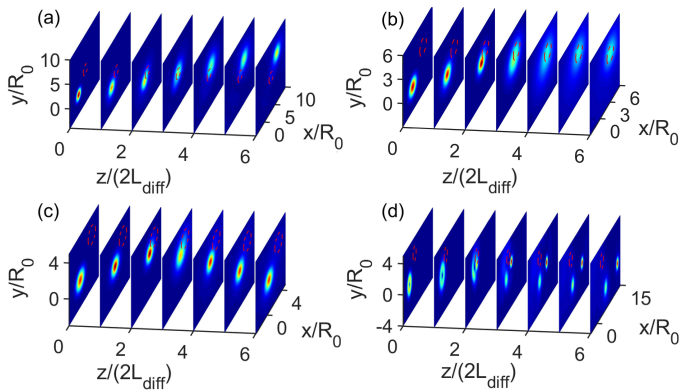}\\
\caption{\footnotesize(Color online)  Scattering of (2+1)D nonlocal optical soliton by a Rydberg defect, by taking the soliton amplitude $|U|$  as a function of $\xi=x/R_0$, $\eta=y/R_0$, and $s=z/(2L_{\rm diff})$.
The soliton (bright spot) is transmitted for $\Omega_{p0}=18\,{\rm MHz}$  [panel (a)], trapped for $\Omega_{p0}=19.2\,{\rm MHz}$ [panel (b)], or reflected for $\Omega_{p0}=20.4\,{\rm MHz}$ [panel (c)] from the defect.
The dashed red circle denotes the central position of the defect.
(d)~The scattering of a (2+1)D nonlocal optical vortex for $\Omega_{p0}=31.3$\,MHz, which disintegrates into two solitons after colliding with the Rydberg defect.}
\label{fig7}
\end{figure}

Shown in Fig.~\ref{fig7}
is the result of the simulation, where panels (a), (b), and (c) are for the cases of $\Omega_{p0}=18\,{\rm MHz}$, $19.2\,{\rm MHz}$, and $20.4\,{\rm MHz}$, respectively. In the figure, the  dashed red circle denotes the central position of the Rydberg defect. We see that, similar to (1+1)D case,
the soliton (illustrated by bright spot) is transmitted for $\Omega_{p0}=18\,{\rm MHz}$  [panel (a)], trapped for $\Omega_{p0}=19.2\,{\rm MHz}$ [panel (b)], and reflected for $\Omega_{p0}=20.4\,{\rm MHz}$ [panel (c)] from the defect.

As an extension, the scattering of (2+1)D nonlocal optical vortices is also investigated. Although nonlocal optical vortices with lower-order angular momenta may be quite stable during free propagation in the Rydberg atomic gas~\cite{Bai2019}, we find that they are generally split into several parts after colliding with the Rydberg defect. Fig.~\ref{fig7}(d) shows the result of the scattering of a vortex taking a Laguerre-Gauss mode with radial index $p=0$ and azimuthal index $m=1$ for $\Omega_{p0}=31.3$\,MHz. We see that the vortex is disintegrated into two solitons after colliding with the Rydberg defect.

\section{Summary}\label{summary}

In this work, we have suggested a scheme to realize a QR of nonlocal nonlinear optical beams in a cold atomic gas via a dispersive Rydberg-EIT. By an approach beyond mean-field approximation, we have derived a nonlinear envelope equation for the propagation of probe laser field, which includes a local linear attractive (Rydberg defect) potential and a nonlocal nonlinear attractive potential. We have demonstrated that the system supports nonlocal optical solitons, which have very low power and display a sharp transition between reflection and transmission when they are scattered by the Rydberg defect potential.
We have found that, different from conventional QRs, the QR of the nonlocal optical solitons in our system depend significantly on the nonlocality degree of the Kerr nonlinearity and display anomalous and controlled QR behaviors contributed by Rydberg blockade effect.

QR is a specific interference phenomenon of matter waves that is absent for the motion of classical particles and general wave motions. Compared to the conventional QR, the QR reported in our study is easy to realize and can be controlled and manipulated actively. Our work open an avenue for studying the QR in systems with nonlocal nonlinearity and for revealing novel optical phenomena based on cold Rydberg atomic gases. The research results reported here are not only useful for developing Rydberg nonlinear optics and characterizing physical property of Rydberg gases, but also helpful for designing novel optical devices at weak-light level.

\section*{Acknowledgments}

We thank T. Pohl, W. Li, and C. Hang for fruitful discussions. This work was supported by the National Natural Science Foundation of China under Grant Nos.~11975098, 11904104, and 11847221, by the China Postdoctoral Science Foundation under Grant No.~2017M620140, by the International Postdoctoral Exchange Fellowship Program under Grant No.~20180040, and by the Shanghai Sailing Program under Grant No.~18YF1407100.

\appendix

\section{Expansion equations of density-matrix elements and their solutions}\label{ap1}

Due to the strong Rydberg-Rydberg interaction, the Bloch equations (\ref{Eq21}) and (\ref{Eq22}) for one-atom density matrix elements
  $\rho_{\alpha\beta}(\mathbf{r},t)\equiv\langle \hat{S}_{\alpha\beta}
(\mathbf{r},t)\rangle$ involve two-atom density-matrix elements
$\rho_{\alpha\beta,\mu\nu}(\mathbf{r}^\prime,\mathbf{r},t)\equiv\langle \hat{S}_{\alpha\beta}
(\mathbf{r}^\prime,t)\hat{S}_{\mu\nu}(\mathbf{r},t)\rangle$,
and hence when solving the one-atom density matrix elements
one must solve the equations of motions for these two-atom density-matrix elements. However, the equations for the two-atom density-matrix elements involve three-atom density-matrix elements, and so on. For large atom density, these equation chains must be solved by using suitable techniques beyond mean-field approximation.

Here we adopt the method developed in Refs.~\cite{Bai,ZhangQi,Hang,Bai2019} to solve these equations under the condition of Rydberg-EIT. We assume that initially all atoms are prepared in the ground state $|1\rangle$. Since the probe field is weak, we can take it as a small parameter (i.e. $\Omega_{p}\sim\epsilon$) to make a perturbation expansion, which reads
$\rho_{\alpha1}=\sum_{l=0}\epsilon^{2l+1}\rho_{\alpha1}^{(2l+1)}$, $\rho_{32}=\sum_{l=1}\epsilon^{2l}\rho_{32}^{(2l)}$, $\rho_{\beta\beta}=\sum_{l=0}\epsilon^{2l}\rho_{\beta\beta}^{(2l)}$
[$\rho_{\beta\beta}^{(0)} =\delta_{\beta1}\delta_{\beta1}$ ($\alpha=2,3; \beta=1,2,3$)],
$\rho_{\alpha\beta,\mu\nu}=\sum_{l=2}\epsilon^{l}\rho_{\alpha\beta,\mu\nu}^{(l)}$.
 Substituting this expansion into equations (\ref{Eq21}), (\ref{Eq22})  and (\ref{Max}), and those for high-order correlators,
and comparing the coefficients of $\epsilon^l$ $(l=1,2,3...)$, we obtain a chain of linear but inhomogeneous equations which can be solved order by order.

\vspace{5mm}
{\sl First-order approximation ($l=1$)}:
The solution in this order describes the linear excitation of the system and no Rydberg-Rydberg interaction is involved. It reads $\rho_{21}^{(1)}=a_{21}^{(1)}\Omega_p$ and $\rho_{31}^{(1)}=a_{31}^{(1)}\Omega_p$, with
\begin{equation}\label{A211}
a_{21}^{(1)}=(d_{31}-\Delta_d)/D_1,
\end{equation}
$a_{31}^{(1)}=-\Omega_c/D_1$, and $D_1=|\Omega_c|^2-d_{21}(d_{31}-\Delta_d)$  [$\Delta_d\equiv \Delta_d(x,y)$ is given by Eq.~(\ref{Delta})]. Other density-matrix elements are zero.

\vspace{5mm}
{\sl Second-order approximation ($l=2$)}: In this order the solution for the diagonal elements reads $\rho_{\alpha\alpha}^{(2)}=a_{\alpha\alpha}^{(2)}|\Omega_p|^2$
($\alpha=1,2,3$), with
\begin{subequations} \label{secondorder}
\begin{eqnarray}
&& a_{11}^{(2)}=\frac{[i\Gamma_{23}-2|\Omega_c|^2 M_1]M_2 -i\Gamma_{12}
|\Omega_c|^2M_3}{-\Gamma_{12}\Gamma_{23}-
i\Gamma_{12}|\Omega_c|^2 M_1},\\
&& a_{33}^{(2)}=\frac{1}{i\Gamma_{12}}\left(M_2-i\Gamma_{12}a_{11}^{(2)}
\right),\\
&& a_{22}^{(2)}=-a_{11}^{(2)}-a_{33}^{(2)},\\
&& a_{32}^{(2)}=\frac{1}{d_{32}}\left(-\frac{\Omega_c}{D_1}
+2\Omega_ca_{33}^{(2)}+\Omega_ca_{11}^{(2)}\right),
\end{eqnarray}
\end{subequations}
with
\begin{subequations} \label{secondorder1}
\begin{eqnarray}
&&M_1=\frac{1}{d_{32}-\Delta_d}-\frac{1}{d_{32}^{\ast}-\Delta_d},\\
&& M_2=\frac{d_{31}^{\ast}-\Delta_d}{D_1}^{\ast}-\frac{d_{31}-\Delta_d}{D_1}, \\
&& M_3=\frac{1}{D_1^{\ast}(d_{32}^{\ast}-\Delta_d)}-\frac{1}{D_1(d_{32}-\Delta_d)}.
\end{eqnarray}
\end{subequations}

The two-atom density-matrix elements $\rho_{\alpha\beta,\mu\nu}$ have non-zero solutions only starting from the second-order approximation. Based on the above results, we can obtain the equations for them, which are given by
\begin{align}\label{twobody1}
&\begin{bmatrix}2d_{21} & 0 & 2\Omega_c^\ast \\
    0 & 2d_{31}-2\Delta_d-V_s & 2\Omega_c \\
    \Omega_c & \Omega_c^\ast & d_{21}+d_{31}-\Delta_d
\end{bmatrix}
\begin{bmatrix}
\rho_{21,21}^{(2)}\\ \rho_{31,31}^{(2)}\\ \rho_{31,21}^{(2)}
\end{bmatrix}\nonumber\\
&=\begin{bmatrix}
-2\frac{d_{31}}{D_1}\\0\\ \frac{\Omega_c}{D_1}
\end{bmatrix}\Omega_p^2,
\end{align}
\begin{align}\label{twobody2}
&\begin{bmatrix}d_{21}+d_{12} & 0 & -\Omega_c & \Omega_c^\ast\\
    -\Omega_c^\ast & \Omega_c^\ast & d_{21}+d_{13}+\Delta_d & 0 \\
    0 & d_{31}+d_{13} & \Omega_c &-\Omega_c^\ast \\
    -\Omega_c & \Omega_c &0 & d_{21}^\ast+d_{13}^\ast+\Delta_d
\end{bmatrix}
\notag\\
&\times
\begin{bmatrix}
\rho_{21,12}^{(2)}\\ \rho_{31,13}^{(2)}\\ \rho_{21,13}^{(2)} \\ \rho_{21,13}^{\ast(2)}
\end{bmatrix}
=\begin{bmatrix}
\frac{d_{31}}{D_1}-\frac{d_{31}^\ast}{D_1^\ast}\\ \frac{\Omega_c^\ast}{D_1^\ast}\\0\\ \frac{\Omega_c}{D_1}
\end{bmatrix}|\Omega_p|^2,
\end{align}
with $\rho_{\alpha1,\beta1}^{(2)}=a_{\alpha1,\beta1}^{(2)}\Omega_p^2$,
$\rho_{\alpha1,1\beta}^{(2)}=a_{\alpha1,1\beta}^{(2)}|\Omega_p|^2$ $(\alpha, \beta=2,3)$.
Expression of $\rho_{\alpha\beta,\mu\nu}^{(2)}$ can be directly obtained by solving Eqs.~(\ref{twobody1}) and (\ref{twobody2}), which are lengthy thus not written down explicitly here.

\vspace{5mm}
{\sl Third-order approximation ($l=3$)}:
Equations for two-atom correlators $\rho_{\alpha\beta,\mu\nu}$ at this order read
\begin{widetext}
\begin{align}\label{C4}
  &\begin{bmatrix}\begin{matrix}M_{31} & \Omega_c^\ast & -i\Gamma_{23} & 0 & \Omega_c^\ast & -\Omega_c & 0 & 0 \\
    \Omega_c & M_{32} & 0 & -i\Gamma_{23} & 0 & 0 & \Omega_c^\ast & -\Omega_c \\
    0 & 0 & M_{33} & \Omega_c^\ast & -\Omega_c^\ast & \Omega_c & 0 & 0 \\
    0 & 0 & \Omega_c & M_{34} & 0 & 0 & -\Omega_c^\ast & \Omega_c \\
    \Omega_c & 0 & -\Omega_c & 0 & M_{35} & 0 & \Omega_c^\ast & 0 \\
    -\Omega_c^\ast & 0 & \Omega_c^\ast & 0 & 0 & M_{36} & 0 & \Omega_c^\ast \\
    0 & \Omega_c & 0 & -\Omega_c & \Omega_c & 0 & M_{37} & 0 \\
    0 & -\Omega_c^\ast & 0 & \Omega_c^\ast & 0 & \Omega_c & 0 & M_{38}
\end{matrix}\end{bmatrix}\begin{bmatrix}
\begin{matrix}\rho_{22,21}^{(3)}\\ \rho_{22,31}^{(3)}\\ \rho_{33,21}^{(3)}\\ \rho_{33,31}^{(3)}\\ \rho_{32,21}^{(3)}\\ \rho_{21,23}^{(3)}\\ \rho_{32,31}^{(3)}\\ \rho_{31,23}^{(3)}
\end{matrix}
\end{bmatrix}
=\begin{bmatrix}
\begin{matrix}-a_{21,12}^{(2)}+a_{21,21}^{(2)}-a_{22}^{(2)}\\
    -a_{31,12}^{(2)}+a_{21,31}^{(2)}\\-a_{33}^{(2)}\\0\\ a_{21,31}^{(2)}-a_{32}^{(2)}\\-a_{32}^{\ast(2)}-a_{21,13}^{(2)}\\ a_{31,31}^{(2)}\\
    -a_{31,13}^{(2)}
\end{matrix}
\end{bmatrix}|\Omega_p({\bf r^\prime})|^2\Omega_p({\bf r}),
\end{align}
\end{widetext}
where $M_{31}=i\Gamma_{12}+d_{21}$, $M_{32}=i\Gamma_{12}+d_{31}-\Delta_d$, $M_{33}=i\Gamma_{23}+d_{21}$, $M_{34}=d_{31}+i\Gamma_{23}-\Delta_d-V_s$, $M_{35}=d_{32}+d_{21}-\Delta_d$, $M_{36}=d_{23}+d_{21}+\Delta_d$, $M_{37}=d_{32}+d_{31}-2\Delta_d-V_s$ and $M_{38}=d_{23}+d_{31}$. From these equations we obtain the third order solution
$\rho_{33,31}^{(3)}=a_{33,31}^{(3)}|\Omega_p({\bf r^\prime})|^2\Omega_p({\bf r})$,
with
\begin{equation}\label{order33313}
\begin{aligned}
a_{33,31}^{(3)}&=\frac{P_0+P_1V_s(\mathbf{r}^\prime-\mathbf{r})
+P_2V_s(\mathbf{r}^\prime-\mathbf{r})^2}{Q_0
+Q_1V_s(\mathbf{r}^\prime-\mathbf{r})+Q_2V_s(\mathbf{r}^\prime-\mathbf{r})^2
+Q_3V_s(\mathbf{r}^\prime-\mathbf{r})^3},
\end{aligned}
\end{equation}
where $D_1=|\Omega_c|^2-d_{21}(d_{31}-\Delta_d)$, $D_2=|\Omega_c|^2-d_{21}(d_{21}+d_{31}-\Delta_d)$, $P_n$ and $Q_n~(n=0,1,2,3)$ are functions of the spontaneous emission decay rate $\gamma_{\mu\nu}$, detunings $\Delta_\mu$, and half Rabi frequency $\Omega_c$.

Consequently, the solution of $\rho_{21}^{(3)}$ takes the form
\begin{eqnarray}\label{}
\rho_{21}^{(3)}=&& a_{21,1}^{(3)}|\Omega_p|^2\Omega_p\nonumber\\
&& +{\cal N}_a\int {\rm d}^3 r^{\prime} a_{21,2}^{(3)}(\mathbf{r},
\mathbf{r}^\prime)|\Omega_p^(\mathbf{r}^\prime)
|^2\Omega_p(\mathbf{r}),
\end{eqnarray}
with the coefficients given by
\begin{eqnarray}\label{rho213}
&& a_{21,1}^{(3)}=\frac{\Omega_c^{\ast}a_{32}^{(2)}+(\omega+d_{31}-\Delta_d)
(2a_{11}^{(2)}+a_{33}^{(2)})}{D_1},\label{A2113}\\
&& a_{21,2}^{(3)}=\frac{-2|\Omega_c|^4(d_{21}+d_{31}-\Delta_d)
V_s(\mathbf{r}^\prime-\mathbf{r})/(|D_1|^2 D_1)}{2d_{21}|\Omega_c|^2+D_2[2d_{31}-2\Delta_d-V_s(\mathbf{r}^\prime-
\mathbf{r})]},\label{A2123}\nonumber\\
\end{eqnarray}
where $D_2=|\Omega_c|^2-d_{21}(d_{21}+d_{31}-\Delta_d)$.

Combining the results given by $\rho_{21}^{(1)}$ and
$\rho_{21}^{(3)}$ given above, we obtain the optical susceptibility for the probe field with the form
\begin{eqnarray}\label{chiapp}
\chi_p
&&= {\cal N}_a ( {\bf e}_p\cdot {\bf p}_{12})^2 \rho_{21}/(\varepsilon_0 \hbar\Omega_{p}),\nonumber\\
&&=\chi_p^{(1)}+\chi_{p,1}^{(3)}|\Omega_p|^2 +\int {\rm d}^3 r^{\prime}\chi_{p,2}^{(3)}(\mathbf{r}^\prime-\mathbf{r})
|\Omega_p(\mathbf{r}^\prime)|^2 ,\nonumber\\
\end{eqnarray}
where $\chi_p^{(1)}={\cal N}_a|{\bf e}_p\cdot\mathbf{p}_{12}|^2a_{21}^{(1)}/(\varepsilon_0\hbar)$ is linear susceptibility, $\chi_{p,1}^{(3)}={\cal N}_a|{\bf e}_p\cdot\mathbf{p}_{12}|^2 a_{21,1}^{(3)}/(\varepsilon_0\hbar)$ is local nonlinear susceptibility, and $\chi_{p,2}^{(3)}={\cal N}_a^2|\mathbf{p}_{12}|^2a_{21,2}^{(3)}/(\varepsilon_0\hbar)$ describes the kernel of  nonlocal nonlinear susceptibility contributed by the long-range Rydberg-Rydberg interaction in the system.



\begin{references}
\bibitem{note3}
The relation between the incident velocity $v$ and the incident angle $\theta$ is given by $\theta={\rm tan}^{-1}[v/(2k_pR_0)]$.
\bibitem{note4}
Fig.~\ref{fig5}(e) is obtained via a numerical simulation by assuming that the incident probe beam has the form $U={\rm sech}(\xi-3)$.
\bibitem{note5}
The repulsion property of the nonlinear potential $V_{\rm non}$
near the Rydberg defect is due to the Rydberg blockade effect contributed by the gate photons stored in the atomic arrays at the defect. Close to the defect, $\Delta_d\rightarrow+\infty$, which suppresses the value of  $W_2\, (\propto\chi_{p,2}^{(3)})$  significantly.
\bibitem{note7}
That is to say, we now make the calculation with the use of 2D defect potential $V=V(\xi,\eta)$ and 2D coefficient of the nonlocal Kerr nonlinearity $W_2=W_2(\xi,\eta)$ in Eq.~(\ref{nonlinearEq}).
\end{references}
\end{document}